\documentclass[conference]{IEEEtran}

\usepackage{cite}
\usepackage{amsmath,amsthm,amssymb,amsfonts}
\usepackage{graphicx}
\usepackage{textcomp}
\usepackage{xcolor}
\usepackage{color}
\usepackage{setspace}

\usepackage{mathrsfs}
\usepackage{pifont}
\usepackage{bm}
\usepackage{pgfplots}
\usepackage{tikz}
\usetikzlibrary{arrows}
\usepackage{subfigure}
\usepackage{graphicx,booktabs,multirow}
\usepackage{upgreek}
\usepackage{bm}
\usepackage{mathrsfs}
\usepackage{algorithm}
\usepackage{algorithmicx}
\usepackage{algpseudocode}
\newcommand{\vct}[1]{{\bm #1}}

\definecolor{colorhkust}{RGB}{20,43,140}
\definecolor{colortsinghua}{RGB}{116,52,129}
\definecolor{color1}{RGB}{128,0,0}

\newtheorem{lemma}{Lemma}
\newtheorem{theorem}{Theorem}

\newtheorem{proposition}{Proposition}
\newtheorem{definition}{Definition}
\newtheorem{remark}{Remark}
\newtheorem{example}{Example}

\newcommand{\tabincell}[2]{\begin{tabular}{@{}#1@{}}#2\end{tabular}}

\newcommand{\trace}{{\rm Tr}}

\mathchardef\re="023C
\mathchardef\im="023D

\IEEEoverridecommandlockouts
\begin{document}

\title{AI Empowered Resource Management for Future Wireless Networks}

%

\author{
   \IEEEauthorblockN{Yifei Shen$^\dagger$, Jun Zhang$^\star$, S.H. Song$^{\dagger}$, and Khaled B. Letaief$^{\dagger\ddagger}$}\\
  \IEEEauthorblockA{$^\dagger$Dept. of ECE, The Hong Kong University of Science and Technology, Hong Kong\\
  	$^\star$ Dept. of EIE, The Hong Kong Polytechnic University, Hong Kong\\
  	$^\ddagger$ Peng Cheng Laboratory, Shenzhen, China\\
  	Email:{ yshenaw@connect.ust.hk,  jun-eie.zhang@polyu.edu.hk, eeshsong@ust.hk, eekhaled@ust.hk} 
  	\thanks{This work was supported by the Hong Kong Research Grants Council under Grant No. 16210719 and 15207220.}}

}

\maketitle

\begin{abstract} Resource management plays a pivotal role in wireless networks, which, unfortunately, leads to challenging NP-hard problems. Artificial Intelligence (AI), especially deep learning techniques, has recently emerged as a disruptive technology to solve such challenging problems in a real-time manner. However, although  promising results have been reported, practical design guidelines and performance guarantees of AI-based approaches are still missing. In this paper, we endeavor to address two fundamental questions: 1) What are the main advantages of AI-based methods compared with classical techniques; and 2) Which neural network should we choose for a given resource management task. For the first question, four advantages are identified and discussed. For the second question, \emph{optimality gap}, i.e., the gap to the optimal performance, is proposed as a measure for selecting model architectures, as well as, for enabling a theoretical comparison between different AI-based approaches. Specifically, for $K$-user interference management problem, we theoretically show that graph neural networks (GNNs) are superior to multi-layer perceptrons (MLPs), and the performance gap between these two methods grows with $\sqrt{K}$.
\end{abstract}


\section{Introduction}\label{sec:intro}
The modern wireless communication industry has experienced several generations of creative development for several decades. Future wireless networks, including 5G networks and beyond will support eMBB (enhanced broadband), uRLLC (ultra-reliable and low-latency communications), and mMTC (massive machine type communications). 
To support such innovative applications, effective large-scale resource management will play a vital role. Unfortunately, typical resource management problems, such as subcarrier allocation, user association, and computation offloading, are non-convex and computationally challenging. Moreover, they need to be solved in a real-time manner in the presence of time varying wireless channels, given the latency requirement of novel mobile applications. Existing algorithms are often based on convex optimization tools, which suffer from sub-optimal performance for non-convex problems and scale poorly with the problem size. 

Motivated by the recent successes of AI techniques, especially deep learning (DL), in computer vision and natural language processing, AI-based methods have been proposed to solve the challenging wireless resource management problems \cite{sun2018learning,lee2018deep,liang2018towards,shen2018lora,nasir2019multi,huang2019deep,shen2020graph,jiang2020learning}. The main purpose is to achieve near-optimal performance in multiple applications including power control \cite{sun2018learning,lee2018deep,liang2018towards,nasir2019multi}, beamforming \cite{shen2018lora,shen2020graph}, computation offloading \cite{huang2019deep}, and intelligent reflection surfaces \cite{jiang2020learning}, in a real-time manner.

Existing methods can be classified into two categories. The first category is  based on a data-driven approach \cite{sun2018learning,lee2018deep,liang2018towards,cui2018spatial}. These methods treat the neural network as a black box and use it to approximate the optimal solution of a given optimization problem. For example, multi-layer perceptrons (MLPs) are adopted to approximate the input-output mapping of the classic weighted minimum mean square error (WMMSE) algorithm to speedup computation \cite{sun2018learning}. Although they can achieve good performance for some specific settings, the black box nature of these methods leads to two major issues, namely, poor interpretability and high dependence on the quality of training data. The second category is a model-driven approach \cite{he2018deep,he2018model,he2019model,hu2020iterative,hu2021joint}, which nicely addresses these two issues by introducing the inductive bias of optimization-based algorithms into neural networks \cite{he2019model}. Specifically, they unroll one iteration of a classic algorithm as one layer of a neural network and replace the ineffective policies in the algorithms by neural networks. However, the unrolled algorithm should be carefully chosen and it often suffer from the model mismatch issue \cite{he2019model}. 

More recently, some intermediate methods have been proposed, which enjoy the benefits of both approaches. For example, the message passing graph neural network (MPGNN) \cite{xu2018powerful,shen2020graph} is a data-driven approach, which can also be viewed as an unrolled decentralized algorithm \cite{shen2020graph}. MPGNNs have shown their superior performance, scalability, and interpretability in the beamformer design \cite{shen2020graph} and phase shifter design \cite{jiang2020learning}  problems. Despite all of these efforts, two fundamental questions from machine learning perspectives remain open:
\begin{enumerate}
    \item What are the main advantages of AI-based methods compared with classical methods?
    \item Which neural network should we use for a specific resource management task?
\end{enumerate}
There have been some attempts to address these questions. Nevertheless, they mainly rely on empirical results. In this paper, we attempt to develop theoretical justifications and practical guidelines. For the first question, we investigate the recent development in nonconvex optimization and wireless communication and identify four unique advantages. For the second question, we refer to the recent development in provably approximate correct (PAC) learning theory. Based on the algorithm alignment framework \cite{xu2019what}, we will show that for the $K$-user interference management problem, graph neural networks (GNNs) are superior to multi-layer perceptrons (MLPs), and the performance gap between these two methods grows with $\sqrt{K}$.

\section{Advantages of AI-based Approaches} \label{sec:advantages}
This section identifies key advantages of AI-enabled resource management algorithms, which can be utilized to identify the proper scenario to apply AI-based methods.

\subsection{Solving NP-hard Problems in Real-time} Many radio resource management problems are NP-hard. This means that there does not exist a polynomial-time algorithm that can obtain the optimal solution. Fortunately, this is the worst-case complexity and in practice we often focus on the average complexity, which allows a tractable algorithm design. For example, the blind data detection problem is NP-hard, but efficient and optimal algorithms can be designed if we consider Rayleigh fading channels \cite{dong2020blind,xue2020blind,shen2020complete}. Despite these positive results, the optimal algorithm should be designed for each individual resource management problem, which requires tremendous efforts. Instead of designing algorithms specialized for each problem, AI-based approach can learn an optimal real-time algorithm from the training data \cite{sun2018learning,lee2018deep,shen2018lora,shen2020graph}. The learnable algorithm will fit the problem automatically. In addition, as neural networks often only involve computationally cheap operations, e.g., matrix multiplication, the real-time constraint can be met in this way.

\subsection{Automatic Design of Distributed Algorithms} In future wireless networks, innovative distributed architectures will be adopted, e.g., cell-free massive MIMO and distributed MIMO systems \cite{interdonato2019ubiquitous}. Thus, it is highly desirable to have effective distributed algorithms for resource management. A good distributed algorithm is extremely difficult to design given practical constraints such as limited backhaul capacity and stringent latency requirements. Fortunately, similar to the discussion in the last subsection, the optimal distributed algorithm can be learned automatically. To meet the distributed requirement, specialized neural network architectures, e.g., GNNs, should be adopted \cite{nasir2019multi,shen2020graph,wang2021decentralized,lee2021decentralized}.

\subsection{Handling Imperfect Measurements} Most resource management algorithms assume perfect channel state information  (CSI), which may not be available in practice. As a result, there have been a line of works dealing with robust resource management with imperfect CSI, for which an AI-based method has its advantage. For example, given an imperfect input, the neural network will first ``calibrate'' it to an accurate one, and then allocate resources according to the calibrated input. It has been shown that without a specialized design for the uncertainty, AI-based methods are already robust to missing CSI \cite{shen2019graph}, noisy CSI \cite{hu2020iterative}, and delayed CSI \cite{nasir2019multi}. Specialized training schemes can be designed to further improve the robustness of AI-based approaches \cite{cui2020deep,sun2020learning,sun2021learning}.
\begin{table*}[htb]
	
	\selectfont  
	\centering
	\caption{Typical training schemes for AI-based methods in wireless communications.} 
	
	\resizebox{0.7\textwidth}{!}{
		\begin{tabular}{|c|c|c|c|c|c|c|}  
			\hline  
			\textbf{Training Schemes} & \textbf{Suitable Conditions} & \textbf{Examples} \\ \hline
			Supervised Learning &  \tabincell{c}{(1) solution is unique and \\(2) a good classic algorithm}
		& \tabincell{c}{Power Control \cite{sun2018learning}, Scheduling \cite{cui2018spatial}, \\User Association \cite{zappone2018user} }
		
		      \\ \hline
			Unsupervised Learning &   \tabincell{c}{(1) solution is non-unique or \\(2) automatic algorithm design}   & (Hybrid) Beamforming \cite{shen2020graph,lin2019beamforming}, IRS \cite{jiang2020learning} \\ \hline
			Reinforcement Learning & \tabincell{c}{(1) long-term planning or 
			\\(2) action affects the states} & \tabincell{c}{ Caching \cite{sadeghi2019deep} \\ Vehicles (UAVs) \cite{ye2018deep,liu2019reinforcement}} \\ \hline
			
	\end{tabular}}
	\label{tab:training}
\end{table*}

\subsection{End-to-end Design} Most existing resource management methods first estimate the channel states and then allocate the radio resources. Such a method has two drawbacks. First, in a large-scale system, CSI estimation introduces non-negligible latency. Second, CSI is estimated with some artificial metrics, e.g., MSE, which may not be optimal to achieve the final resource management goal. With AI-based resource management, the two stages can be unified in an end-to-end manner, i.e., the neural network directly allocates resources based on the received pilots without the need to explicitly estimate CSI. This can significantly reduce the pilot overhead and improve performance \cite{sohrabi2021deep,jiang2020learning,guo2020deep,ma2021neural}.

Note that the first advantage has been well identified in the literature of operation research \cite{bengio2018machine} while the latter three are unique in wireless networks.

\section{Comparisons among Training Schemes}
In the following two sections, we focus on the second question. The two key components of AI-based resource management are neural network architectures and the training schemes, which are \emph{independent} from each other. In this section, we discuss the practical guidelines for the selection of training schemes and we will investigate the selection of neural network architectures in the next section. The training scheme often involves three categories: supervised, unsupervised, and reinforcement learning. We will discuss them in the sequel.

\subsection{Supervised Learning}
 With supervised learning, the resource management problem is first transformed into a classification or regression problem, and then loss functions are adopted from statistical machine learning. For example, the power control problem can be regarded as a regression problem \cite{sun2018learning}, and the wireless scheduling problem can be considered as a binary classification problem \cite{cui2018spatial}. Due to its simplicity, supervised learning is often adopted in AI-based resource management problems \cite{sun2018learning,shen2018lora,cui2018spatial}. However, when the optimal solution is not unique, they suffer from loss mismatch and perform poorly. To illustrate this issue, we consider the following eigenvector problem, where $\bm{R} \in \mathbb{C}^{n \times n}$, $\bm{v} \in \mathbb{C}^n$
\begin{equation}\label{eq:eigen}
\begin{aligned}
&\underset{\vct{v} \in \mathbb{C}^n }{\text{maximize}}
& & \bm{v}^H \bm{R} \bm{v} \\
& \text{subject to}
& & \|\bm{v}\|_2 \leq 1.
\end{aligned}
\end{equation}
This problem has its practical application in single user beamforming problem \cite{yu2019miso}. Note that if $\bm{v}^*$ is an optimal solution for \eqref{eq:eigen}, then $-\bm{v}^*$ is also an optimal solution to \eqref{eq:eigen}. Denote the output of the neural network as $\hat{v}$, during the training. If we adopt supervised learning such as  \cite{sun2018learning}, the following objective need to be minimized
\begin{align*}
    \ell_{\text{MSE}}(\Theta) = \|\hat{\bm{v} }(\Theta) - \bm{v}^*\|_2^2 + \|\hat{ \bm{v} }(\Theta) - (-\bm{v}^*)\|_2^2,
\end{align*}
where $\Theta$ denotes the learnable parameters in the neural networks. The optimal solution is $\hat{v}(\Theta) \equiv 0$, which results in bad performance.

\subsection{Unsupervised Learning} To address the above issue, one can adopt unsupervised learning, where the objective function in the resource management problem is adopted as the loss function while training the neural network. Besides overcoming this limitation of supervised learning, another advantage is that it does not require training labels. However, as the resource management problem is often non-convex, the neural network is also non-convex, and as such the optimization landscape becomes highly complicated. Due to the difficulty of optimization, unsupervised training does not always outperform the supervised one (see the comparison of Table XIII and XIV in \cite{lee2019graph}). A very recent paper also conducts a comprehensive theoretical comparison between supervised and unsupervised model \cite{song2021supervise}.

\subsection{Reinforcement Learning} Reinforcement learning differs from the above two methods in three aspects. First, the above two methods require the objective function to be (sub)differentiable while reinforcement learning does not. Second, supervised or unsupervised learning often considers a short-term objective while reinforcement learning considers a long-term reward. Third, with reinforcement learning, the resource management result may affect the environment. For example, in resource management for UAV communication networks, the UAV trajectory affects the channel. Under such a circumstance, the dynamically data collection and training property of reinforcement learning will benefit. However, as reinforcement learning does not effectively exploit the first-order information of the reward in training neural networks, it suffers from poor convergence speed and often converges to a bad local minima during training. Thus, it is often outperformed by supervised learning when the solution is unique and there is no loss mismatch issue (see Table in \cite{li2018combinatorial}).

\subsection{Guidelines for the Training Method Selection} We conclude the discussion about the training methods by providing practical guidelines. When there exists a classic algorithm to generate the optimal solution and the solution is unique, supervised learning is more prefered. When the optimal solution is not unique, supervised learning will fail and we should adopt unsupervised learning. When we consider a long-term resource management problem where the resource management results influence the environment, reinforcement learning will offer a good solution. We list typical problems for the three training methods in Table \ref{tab:training}.

\section{Comparisons among Neural Network Architectures} \label{sec:architectures}
So far, we have discussed the benefits of AI-based approaches and the selection of training schemes. Another important aspect is which neural network architecture should we adopt for AI-based resource management. In this section, we will propose optimality gap as a comparison measure and provide examples of comparing different approaches theoretically. 

\subsection{Introduction to Neural Network Architectures}
The widely adopted neural network architectures can be grouped into two categories. The first category is inherited from image and language processing, e.g., MLPs and convolutional neural networks (CNNs). They have been introduced in \cite{zappone2019wireless}, and hence we will not cover them in this subsection. The second category includes architectures considering the unique properties of wireless networks. Specifically, we consider MPGNNs  \cite{shen2020graph}, which exploit the  wireless network topology, and unrolled networks \cite{he2019model}, which leverage the unique objectives such as capacity. 

\paragraph{MPGNNs} MPGNNs are a class of neural networks that operate on graphs, mimicking distributed message passing algorithms in networks \cite{angluin1980local}. In each layer of an MPGNN, the nodes update their representations by aggregating features from the neighbors. Specifically, the update rule of the $k$-th layer at node $i\in\mathcal{V}$ in an MPGNN \cite{xu2018powerful} is
\begin{equation}\label{eq:mpgnn}
\begin{aligned}
 \bm{x}_i^{(k)} = \alpha^{(k)}\left(\bm{x}_i^{(k-1)}, \phi^{(k)} \left(\left\{\left[\bm{x}_j^{(k-1)},\bm{e}_{j,i}\right]: j \in \mathcal{N}(i)   \right\} \right)  \right),
\end{aligned}
\end{equation}
where $\bm{x}_i^{(k)}$ denotes the feature vector of node $i$ at the $k$-th layer, $e_{i,j}$ is the edge feature between node $i$ and $j$, $\mathcal{N}(i)$ is the set of neighbors of $i$, $\phi^{(k)}$ is a learnable aggregation function of node $i$ that collects information from the neighboring nodes, and $\psi^{k}$ is a learnable function that combines the aggregated information with node $i$'s own information. GNNs in the form of \eqref{eq:mpgnn} can be implemented efficiently with Pytorch Geometric \cite{fey2019fast}.

To develop GNN-based resource management algorithms, we first model the wireless networks as graphs. Specifically, we model the agents (e.g., users, base stations, and phase shifters) as nodes of the graph, the communication links as the edges, and the channel states as the edge features. We then paramterize functions $\alpha^{(k)}$ and $\phi^{(k)}$. One popular architecture is to adopt MLPs for $\alpha^{(k)}$ and $\phi^{(k)}$, i.e., 
\begin{equation}\label{eq:cgcnet}
\small
\begin{aligned}
\bm{y}_i^{(k)} &= \text{MLP2}\left(\bm{x}_i^{(k-1)},  \text{MAX}_{j \in \mathcal{N}(i)}\left\{\text{MLP1} \left(\bm{x}_j^{(k-1)},\bm{e}_{j,i}\right)\right\}    \right),\\
\bm{x}_i^{(k)} &= \beta\left(\bm{y}_i^{(k)}\right),
\end{aligned}
\end{equation}
where $\text{MLP1}$ and $\text{MLP2}$ are two different MLPs, $\beta$ is a differentiable normalization function, e.g., power normalization, $\bm{y}_i^{(k)}$ denotes the output of MLP2 at the $i$-th node in the $k$-th layer, and $\bm{x}_i^{(k)}$ denotes the hidden state. This architecture showed superior performance in power control, beamforming, and phase shifter design \cite{shen2020graph,jiang2020learning}.  

\paragraph{Unrolled Networks} Unrolled networks introduce the inductive bias of classic algorithms into  deep learning. The basic idea is to view one iteration of a classic algorithm as one layer of the neural network, and learn the hyperparameters in the algorithm via back propagation. For example, denoting $\bm{H}$ as the channel matrix and  $x^{(k)}$ as the recovered signal at the $k$-th iteration, then the updates of the OAMP (orthogonal approximate message passing) algorithm for MIMO detection can be given by \cite{he2018deep}
\begin{align*}
    &\bm{r}^{(k)} = \bm{x}^{(k)} - \gamma^{(k)} \bm{W}^{(k)} (\bm{y} - \bm{H}\bm{x}^{(k)}), \\
    &\bm{x}^{(k+1)} = \mathbb{E} (\bm{x}| \bm{r}^{(k)}, \tau^{(k)}), \\
    &(v^{(k)})^2 = \frac{\|\bm{y} - \bm{H}\bm{x}^{(k)}\|_2^2 - M\sigma^2}{\trace (\bm{H}^T\bm{H})}, \\
    &(\tau^{(k)})^2 = \frac{1}{2N} \trace(\bm{C}\bm{C}^T)(v^{(k)})^2 + \frac{(\theta^{(k)})^2\sigma^2}{4N} \trace (\bm{C}\bm{C}^T), 
\end{align*}
where $\bm{W}$, $\bm{C}$ are some transformations of the input channel matrix $\bm{H}$. $\bm{r}^{(k)}$, $v^{(k)}$, $\tau^{(k)}$ are intermediate variables at iteration $k$, and $\theta^{(k)}$, $\gamma^{(k)}$ are hyperparameters required to tune. 

The unrolled OAMP algorithm is first implemented with a deep learning toolbox, where the hyperparameters $\gamma^{(k)}$, and $\theta^{(k)}$ are set as learnable parameters. The neural network can be trained in an end-to-end manner by optimizing the distance between the output and the true symbol. In this way, the hyperparameters are optimally tuned for a certain distribution. This architecture is widely adopted in MIMO detection \cite{he2018model}, channel estimation \cite{he2018deep} and precoding \cite{hu2020iterative}.

We will leave the discussion for the best suitable architecture to Section \ref{sec:architectures}.

\subsection{Comparison from Approximation Perspective} \label{sec:sys_model}
In the content of wireless communications, a common approach for comparing different architectures is to investigate the universal approximation property of neural networks \cite{hornik1989multilayer}, which was adopted in \cite{sun2018learning,qiang2019deep,shen2020graph,zheng2021online}. However, such theoretical results fall short in many aspects. First, besides neural networks, there are many other universal approximators, e.g., Fourier series. The arguments in \cite{sun2018learning,qiang2019deep,zheng2021online} cannot distinguish which one is better, not to mention the comparison of two neural network architectures. Second, universal approximation only states that there exists a neural network to have near-optimal training performance, but reveals nothing about the testing performance. In practice, we care about the optimality gap in the test stage. A framework to characterize the optimality gap is needed.



\subsection{Optimality Gap as a Measure} 
In this subsection, we introduce the PAC-learning framework, and show that with AI-based wireless resource management, a high sample efficiency is \emph{equivalent} to a small optimality gap.

The optimality gap, i.e., the gap between the \emph{optimal} objective value and the value achieved by a given method, consists of three terms: approximation gap, training gap, and generalization gap. Let $\mathcal{E}_{\text{gap}}$ denote the optimality gap, then we have

\begin{align*}
    \mathcal{E}_{\text{gap} } \leq \mathcal{E}_{\text{approximation} } + \mathcal{E}_{\text{training} } +  \mathcal{E}_{\text{generalization} }.
\end{align*}
These three terms are discussed in the following.

If the adopted model is a universal approximator, the approximation gap is $0$; otherwise, there is a nonzero approximation gap measured by
\begin{align*}
    \mathcal{E}_{\text{approximation} } = \min_{f \in \mathcal{F}} \mathbb{E}|R(g(x)) - R(f(x))|,
\end{align*}
where $R(\cdot)$ is some metric such as sum capacity, $g(\cdot)$ is an oracle function such that the correct inference result is $y_i = g(x_i)$, and $\mathcal{F}$ is the hypothesis class. For example, MLPs are universal approximators. Hence, their approximation gap is zero, while the unrolled networks are not universal approximators and they have this gap.

The training gap is caused by the training method. After performing a stochastic gradient descent for a limited number of epochs, we may not find the best weights. For example, the training gap of over-parameterized neural networks is given by the following theorem.
\begin{theorem} \cite{du2018gradient} \label{thm:conv} (Training Error $\mathcal{E}_{\text{train}}$ of neural networks)
    Assume the samples $\bm{x}_i \in \mathbb{R}^n$ and labels $y_i$ are bounded. Then if we set the number of hidden nodes as $m = \Omega(\frac{n^6}{\delta^3})$ and we denote $u(t)$ as the output of the neural network at the $t$-th training iteration, for any $\delta$, then we have
        \begin{align*}
            \|\bm{u}(t) - \bm{y}\|_2^2 \leq \exp(-\lambda_0 t) \|\bm{u}(0) - \bm{y}\|_2^2,
        \end{align*}
    with probability at least $1-\delta$ and $\lambda_0$ is the minimal eigenvalue of data gram matrix and $\lambda_0 \in (0,1)$. 
\end{theorem}

\begin{remark}
   From the above discussion, we could achieve a zero approximation gap and training gap if we have 1) an overparameterized neural network; and 2) it has been trained for a sufficient number of epochs.
\end{remark}

Next we discuss the generalization gap, which is formally defined in the PAC framework. 

\begin{definition} (PAC Learning)\cite{valiant1984theory,xu2019what} Fix an error parameter $\epsilon > 0$ and failure probability $\delta \in (0,1)$. Suppose $\{x_i,y_i\}_{i=1}^M$ are i.i.d. samples drawn from distribution $\mathcal{D}$, and the data satisfies $y_i = g(\bm{x}_i)$. Let $f=\mathcal{A}(\{x_i,y_i\}_{i=1}^M)$ be the function generated by a learning algorithm $\mathcal{A}$. Then $g$ is $(M, \epsilon, \delta)$-learnable with $\mathcal{A}$ if 
\begin{align*}
    \mathbb{P} [|R(f(x)) - R(g(x))| \leq \epsilon ] \geq 1 - \delta.
\end{align*}
\end{definition}

The next lemma shows the generalization gap of neural networks.

\begin{lemma} \cite{arora2019fine,xu2019what}
    Let $\mathcal{A}$ be an over-paramterized and randomly initialized two-layer MLP trained with gradient descent for a sufficient number of iterations. Suppose $y_i = g(\bm{x}_i) = \sum_j \alpha_j(\bm{\beta}_j^T \bm{x}_i)^{p_j}$ with $p_j = 1$ or $p_j = 2l$, $\alpha_j \in \mathbb{R}$, $\bm{\beta}_j \in \mathbb{R}^n$. The sample complexity, i.e., the required number of samples, is 
    \begin{align*}
        \mathcal{O}\left(\frac{\sum_{j} p_j |\alpha_j| \|\bm{\beta}\|_2^p}{\epsilon^2} + \log(1/\delta) \right).
    \end{align*}
\end{lemma}
Based on the above results of the three gap terms, the overall optimality gap is characterized in the next proposition.

\begin{proposition} (Optimality Gap of Neural Networks) \label{pro:high}
    Let $\mathcal{A}$ be an over-paramterized and randomly initialized two-layer MLP trained with gradient descent for $t$ iterations on $M$ training samples. Suppose $y_i = g(\bm{x}_i) = \sum_j \alpha_j(\bm{\beta}_j^T \bm{x}_i)$ with $p_j = 1$ or $p_j = 2l$. Then,
    \begin{align*}
        \mathcal{E}_{\text{gap}} = \mathcal{O}\left(\exp(-\lambda_0 t) + \frac{\sum_{j} p_j |\alpha_j| \|\bm{\beta}\|_2^p}{\sqrt{M} } \right).
    \end{align*}
\end{proposition}

\begin{remark}
As neural networks are universal approximators, we do not have an approximation gap in this bound. The first term indicates the training gap. If the neural network is not trained properly, e.g., trained with only a small number of iterations or the data is not normalized, this gap will be large. The second term is the generalization gap. With a finite number of training samples, the gap is smaller if the target function is a simple one. For example, in the beamforming task, maximum ratio transmission (MRT) is simpler than zero forcing (ZF). Thus, a simple MLP can learn MRT well while learning ZF requires more complicated neural networks like GNNs.
\end{remark}

An important message from Proposition \ref{pro:high} is that for neural networks, the optimality gap is inversely propositional to the number of training samples. Thus, a high sample efficiency implies a small optimality gap.

\subsection{Comparison of Neural Network Architectures}
In this subsection, we introduce the algorithm alignment framework proposed in \cite{xu2019what}, and then adopt this framework to compare different neural network architectures for radio resource management. 

\begin{definition} (Algorithmic Alignment) \cite{xu2019what}
    Let $g$ be the oracle function such that $y_i = g(x_i)$, and $\mathcal{N}$ be a neural network with $n$ modules $\mathcal{N}_i$. The module functions $f_1,\cdots, f_n$ generate $g$ for $\mathcal{N}$ if by replacing $\mathcal{N}_i$ with $f_i$, the network $\mathcal{N}$ simulates $g$. Then, the network $\mathcal{N}(M,\epsilon,\delta)$-algorithmically aligns with $g$ if (1) $f_1,\cdots, f_n$ generate $g$ and (2) there are learning algorithms $\mathcal{A}_i$ for the $\mathcal{N}_i$'s such that the total sample complexity is less than $M$.
\end{definition}

The next theorem shows that the algorithmic alignment improves the sample complexity. By the argument in the last subsection, it also improves the optimality gap. 
\begin{theorem} \cite{xu2019what} \label{thm:aa} Fix $\epsilon$ and $\delta$. Suppose $y_i = g(x_i)$ for some $g$. Suppose $\mathcal{N}_1, \cdots, \mathcal{N}_n$ are network $\mathcal{N}$'s MLP modules in the sequential order. Suppose $\mathcal{N}$ and $g$ $(M,\epsilon,\delta)$-algorithmically align via functions $f_1,\cdots, f_n$. Under the same assumptions as in \cite{xu2019what}, $g$ is $(M, O(\epsilon),O(\delta))$-learnable by $\mathcal{N}$.
\end{theorem}

\begin{example} (MLPs versus MPGNNs for Power Control) \label{exp:power}
We give an example of power control to elaborate the power of Theorem \ref{thm:aa} in characterizing the optimality gap. The problem formulation follows Section V.A in \cite{shen2020graph} with $K$ users. In this example, we consider to approximate an oracle distributed local algorithm, namely, the \emph{oracle algorithm} (Algorithm I in \cite{shen2020graph}) with MPGNNs and MLPs. A giant MLP learns the same function $h$ and $g$ repeatedly for $K$ times and encode them in weights. This leads to $\mathcal{O}(K)$ sample complexity reduction according to Theorem \cite{xu2019what}. Based on Proposition \ref{pro:high}, the optimality gap of MLPs is $\mathcal{O}(\sqrt{K}\epsilon)$ if we denote the optimality gap of MPGNNs as $\mathcal{O}(\epsilon)$. Thus, there is a difference of $\sqrt{K}$. This is verified by the experiments in Table I of \cite{shen2020graph}. 
\end{example}

We next discuss more general cases, to unify classic algorithms and deep learning-based approaches in one framework, using optimality gap as a measure. The classic algorithms do not have the training gap and generalization gap as there are no learnable parameters. Nevertheless, the approximation gap is large due to the model mismatch issue. 

To reduce the approximation gap, data-driven approaches were proposed. The very first works employed MLPs \cite{sun2018learning,liang2018towards}. As MLPs are unstructured, it is difficult to train and the generalization gap is large. To improve the generalization, the structures of wireless resource management problems were exploited in \cite{shen2020graph} via adopting MPGNNs. In Example \ref{exp:power}, we have proved that this improves the optimality gap by a factor of $\sqrt{K}$. These architectures are universal approximators, so they have a zero approximation gap and small training gap. Another line of works are model-driven approaches, i.e., unrolled neural networks. They are not universal approximators so they still have an approximation gap. Meanwhile, the generalization gap is small as the number of parameters is often small. 

\begin{figure}[htb]
	\centering
	\includegraphics[width=0.48\textwidth]{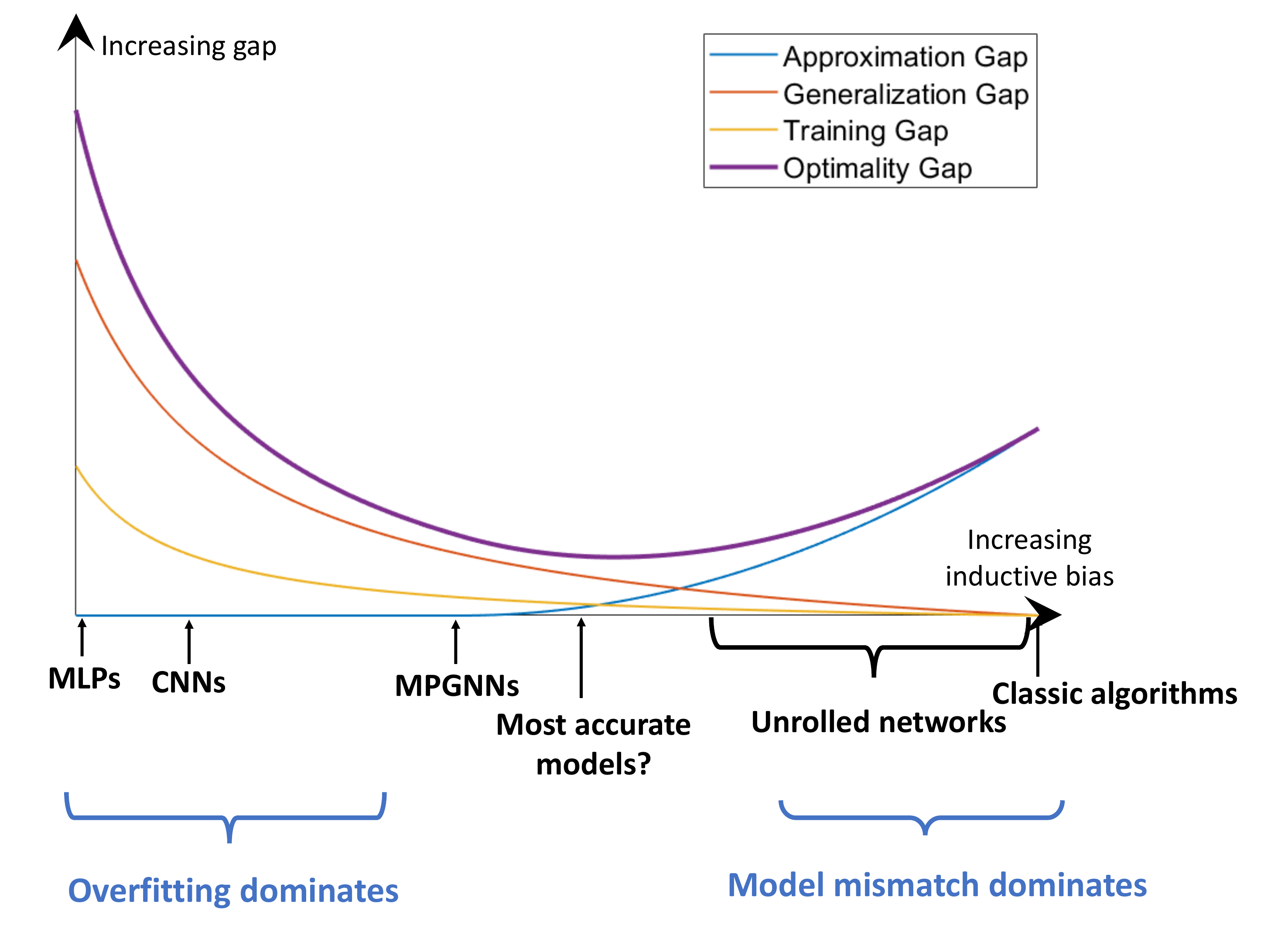}
	\caption{Optimality gap versus inductive bias in different neural network architectures.}
	\label{fig:gap}
\end{figure}

There are some interesting trends in the historical development of different methods. The data-driven approaches started from the unstructured ones (e.g., MLPs \cite{sun2018learning}) and then transited to the structured ones (e.g., MPGNNs \cite{shen2020graph}). For the model driven approaches, the degree of freedom keeps increasing, from tens of parameters in \cite{he2018deep} to thousands of paramters in \cite{hu2020iterative,hu2021joint}. Although these works have made significant efforts, the best model for the wireless resource management is still open. An illustration of the optimality gap for different neural network architectures is shown in Fig. \ref{fig:gap}.

\section{Conclusions}
In this paper, we carried out systematic investigations on AI-based methods for radio resource management. Advantages of such methods were firstly identified, followed by comparisons of different neural network architectures and training schemes, as well as, a theoretical analysis on the generalization performance. Along the discussion, practical design guidelines were provided. While the study is far from complete, the discussions of this paper shed new lights on developing and adopting AI-based methods for radio resource management in future wireless networks \cite{letaief2019roadmap}.

\bibliographystyle{ieeetr}
\bibliography{ref}

\end{document}